\documentclass[10pt, a4paper]{article}
\usepackage{lrec}
\usepackage{xcolor}
\usepackage{graphicx}
\usepackage{tabularx}
\usepackage{soul}
\usepackage{multirow}
\usepackage{caption,subcaption}

\usepackage{epstopdf}
\usepackage[utf8]{inputenc}

\usepackage{hyperref}
\usepackage{xstring}

\title{HotelRec: a Novel Very Large-Scale Hotel Recommendation Dataset}

\name{Diego Antognini, Boi Faltings}

\address{Artificial Intelligence Laboratory \\
  École Polytechnique Fédérale de Lausanne \\
  Lausanne, Switzerland \\
  \texttt{firstname.lastname@epfl.ch}\\}

\abstract{Today, recommender systems are an inevitable part of everyone's daily digital routine and are present on most internet platforms.
State-of-the-art deep learning-based models require a large number of data to achieve their best performance.
Many datasets fulfilling this criterion have been proposed for multiple domains, such as Amazon products, restaurants, or beers.
However, works and datasets in the hotel domain are limited: the largest hotel review dataset is below the million samples.
Additionally, the hotel domain suffers from a higher data sparsity than traditional recommendation datasets and therefore, traditional collaborative-filtering approaches cannot be applied to such data.
In this paper, we propose HotelRec, a very large-scale hotel recommendation dataset, based on TripAdvisor, containing $50$ million reviews.
To the best of our knowledge, HotelRec is the largest publicly available dataset in the hotel domain ($50M$ versus $0.9M$) and additionally, the largest recommendation dataset in a \textit{single domain} and with \textit{textual reviews}~($50M$~versus~$22M$).
    We release HotelRec for further research:  \url{https://github.com/Diego999/HotelRec}. \\ \newline  }

\begin{document}

\maketitleabstract

\section{Introduction}

The increasing flood of information on the web creates a need for selecting content according to the end user's preferences. Today, recommender systems are deployed on most internet platforms and play an important role in everybody's daily digital routine, including e-commerce websites, social networks, music streaming, or hotel booking. 

Recommender systems have been investigated over more than thirty years~\cite{bobadilla2013recommender}. Over the years, many models and datasets in different domains and various sizes have been developed: movies~\cite{harper2016movielens}, Amazon products~\cite{mcauley2015image,he2016ups}, or music~\cite{Celma:Springer2010}. With the tremendous success of large deep learning-based recommender systems, in better capturing user-item interactions, the recommendation quality has been significantly improved~\cite{covington2016deep}.

However, the increase in recommendation performance with deep learning-based models comes at the cost of large datasets. Most recent state-of-the-art models, such as \cite{Wang2019NGC33311}, \cite{liang2018variational}, or \cite{he2017neural} necessitate large datasets (i.e., millions) to achieve high performance.

In the hotel domain, only a few works have studied hotel recommendation, such as \cite{wang2011latent} or \cite{zhang2015hotel}. Additionally, to the best of our knowledge, the largest publicly available hotel review dataset contains $870k$ samples~\cite{li2016understanding}.
Unlike commonly used recommendation datasets, the hotel domain suffers from higher data sparsity and therefore, traditional collaborative-filtering approaches cannot be applied \cite{zhang2015hotel,khaleghi2018comparative,musat2015personalizing}. Furthermore, rating a hotel is different than traditional products, because the whole experience lasts longer, and there are more facets to review ~\cite{khaleghi2018comparative}.
\begin{figure}
    \centering
    \includegraphics[width=\linewidth]{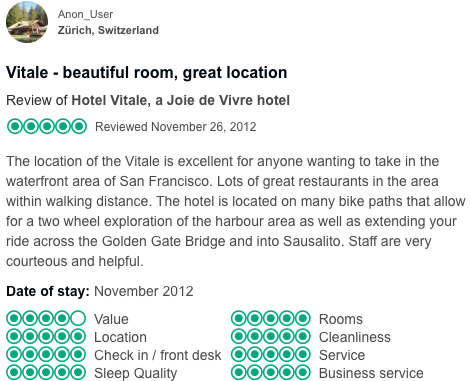}
    \caption{\label{fig_review}Review from TripAdvisor, with sub-ratings. }
    
\end{figure}

\begin{table}[t]
  \centering
  \begin{tabular}{c@{\hspace{1mm}}c@{\hspace{2mm}}c@{\hspace{2mm}}c@{\hspace{1mm}}c}
    HotelRec & \#Users & \#Items & \#Interactions & Sparsity \\
        \hline
        Full & $21\,891\,294$ & $365\,056$ & $50\,264\,531$ &    $99.99937\%$\\
        $5$-core & $2\,012\,162$ & $312\,081$ & $21\,108\,245$ & $99.99664\%$\\
        $20$-core & $72\,603$ & $38\,903$ & $2\,222\,373$ &      $99.92132\%$\\
    \end{tabular}
  \caption{\label{stat_core_dataset}Statistics of the whole HotelRec dataset and its $k$-core subsets (number of users, items, interactions, and the sparsity ratio).}
\end{table}

In contrast, we propose in this work HotelRec, a novel large-scale hotel recommendation dataset based on hotel reviews from TripAdvisor, and containing approximately $50$~million reviews. A sample review is shown in Figure~\ref{fig_review}. To the best of our knowledge, HotelRec is the largest publicly available hotel review dataset (at least $60$~times larger than previous datasets). Furthermore, we analyze various aspects of the HotelRec dataset and benchmark the performance of different models on two tasks: rating prediction and recommendation performance. Although reasonable performance is achieved by a state-of-the-art method, there is still room for improvement. We believe that HotelRec will offer opportunities to apply and develop new large recommender systems, and push furthermore the recommendation for hotels, which differs from traditional datasets. 

\section{Related Work}

Recommendation is an old problem that has been studied from a wide range of areas, such as Amazon products~\cite{mcauley2013hidden}, beers~\cite{McAuley2012}, restaurants\footnote{\url{https://www.yelp.com/dataset/challenge}}, images~\cite{geng2015learning}, music~\cite{Celma:Springer2010}, and movies~\cite{harper2016movielens}. The size of the datasets generally varies from hundreds of thousands to tens of millions of user-item interactions; an interaction always contains a rating and could have additional attributes, such as a user-written text, sub-ratings, the date, or whether the review was helpful. At the time of writing, and to the best of our knowledge, the largest available recommendation corpus on a specific domain and with textual reviews, is based on Amazon Books and proposed by \newcite{he2016ups}. It contains a total of $22$ million book reviews. In comparison, HotelRec has $2.3$~times more reviews and is based on hotels. Consequently, HotelRec is the largest domain-specific public recommendation dataset \textit{with textual reviews} and \textit{on a single domain}. We highlight \textit{with textual reviews}, because some other datasets (e.g., Netflix Prize~\cite{bennett2007netflix}) contain more interactions, that \textit{only} includes the rating and the date.

To the best of our knowledge, only a few number of datasets for hotel reviews have been created: $35$k~\cite{wang2011latent}, $68$k~\cite{musat2013recommendation}, $140$k~\cite{antognini2019multidimensional}, $142$k~\cite{cozza2018mining}, $235$k~\cite{wang2011latent},  $435$k~\cite{musat2015personalizing}, and $870$k~\cite{li2016understanding}. However, the number of users, items, and interactions is limited compared to traditional recommendation datasets. In contrast, the HotelRec dataset has at least two orders of magnitude more examples. Statistics of HotelRec is available in Table~\ref{stat_core_dataset}.

\section{HotelRec}

Everyday a large number of people write hotel reviews on on-line platforms (e.g., Booking\footnote{\url{https://www.booking.com/}}, TripAdvisor\footnote{\url{https://www.tripadvisor.com/}}) to share their opinions toward multiple aspects, such as their \textit{Overall} experience, the \textit{Service}, or the \textit{Location}. Among the most popular platforms, we selected TripAdvisor: according to their third quarterly report of November $2019$\footnote{\url{https://www.sec.gov/ix?doc=/Archives/edgar/data/1526520/000156459019041094/trip-10q_20190930.htm}}, on the \textit{U.S. Securities and Exchange Commission} website\footnote{\url{https://www.sec.gov}}, TripAdvisor is the world's largest online travel site with approximately $1.4$~million hotels. Consequently, we created our dataset HotelRec based on TripAdvisor hotel reviews. The statistics of the HotelRec dataset, the $5$-core, and $20$-core versions are shown in Table~\ref{stat_core_dataset}; each contains at least $k$~reviews for each user or item.

In this section, we first discuss about the data collection process (Section~\ref{sec_data_coll}), followed by general descriptive statistics (Section~\ref{sec_descr}). Finally, Section~\ref{sec_asp_corr} analyzes the overall rating and sub-ratings.

\subsection{Data Collection}
\label{sec_data_coll}
We first crawled all areas listed on TripAdvisor's SiteIndex\footnote{\url{https://www.tripadvisor.com/SiteIndex}}. Each area link leads to another page containing different information, such as a list of accommodations, or restaurants; we gathered all links corresponding to hotels. Our robot then opened each of the hotel links and filtered out hotels without any review. In total, in July $2019$, there were $365\,056$ out of $2\,502\,140$ hotels with at least one review.

Although the pagination of reviews for each hotel is accessible via a URL, the automatic scraping is discouraged: loading a page takes approximately one second, some pop-ups might appear randomly, and the robot will be eventually blocked because of its speed. We circumvented all these methods by mimicking a human behavior with the program \textit{Selenium}\footnote{\url{https://selenium.dev/}}, that we have linked with \textit{Python}\footnote{Using the package \url{https://github.com/SeleniumHQ/selenium/}}. However, each action (i.e., disabling the calendar, going to the next page of reviews) had to be separated by a time gap of one second. Moreover, each hotel employed a review pagination system displaying only five reviews at the same time, which majorly slowed down the crawling.

An example review is shown in Figure~\ref{fig_review}. For each review, we collected: the URL of the user's profile and hotel, the date, the overall rating, the summary (i.e., the title of the review), the written text, and the multiple sub-ratings when provided. These sub-ratings correspond to a fine-grained evaluation of a specific aspect, such as \textit{Service}, \textit{Cleanliness}, or \textit{Location}. The full list of fine-grained aspects is available in Figure~\ref{fig_review}, and their correlation in Section~\ref{sec_asp_corr}

We naively parallelized the crawling on approximately $100$~cores for two months. After removing duplicated reviews, as in \newcite{mcauley2013hidden}, we finally collected $50\,264\,531$ hotel reviews.

\subsection{Descriptive Statistics}
\label{sec_descr}

\begin{figure*}[!ht]
    \centering 
\begin{subfigure}{0.45\textwidth}
  \includegraphics[width=\linewidth]{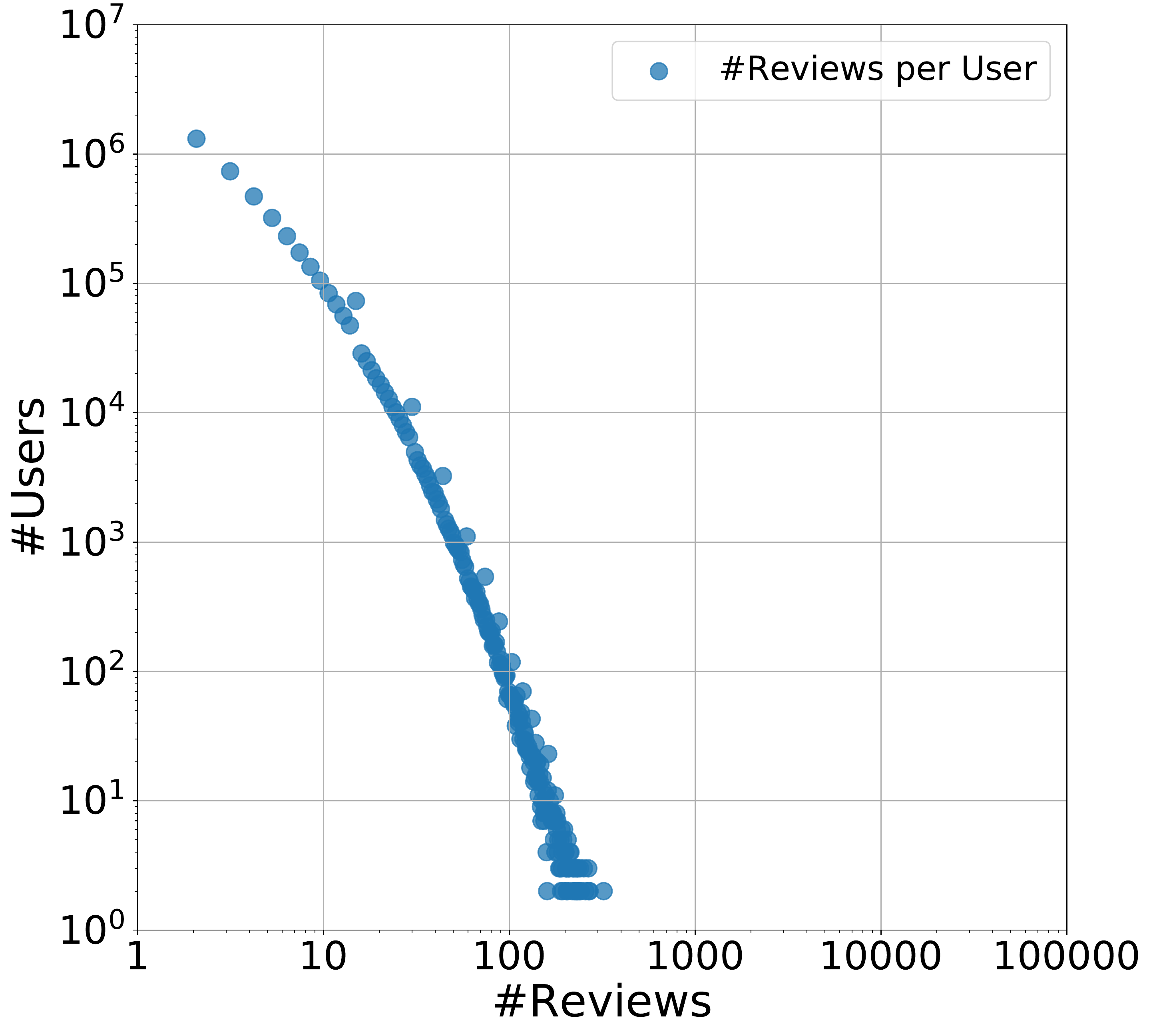}
  \caption{\label{fig_nb_reviews_user}Number of reviews per user. Mean: $2.24 \pm 3.95$.}
\end{subfigure}\hfil 
\begin{subfigure}{0.45\textwidth}
  \includegraphics[width=\linewidth]{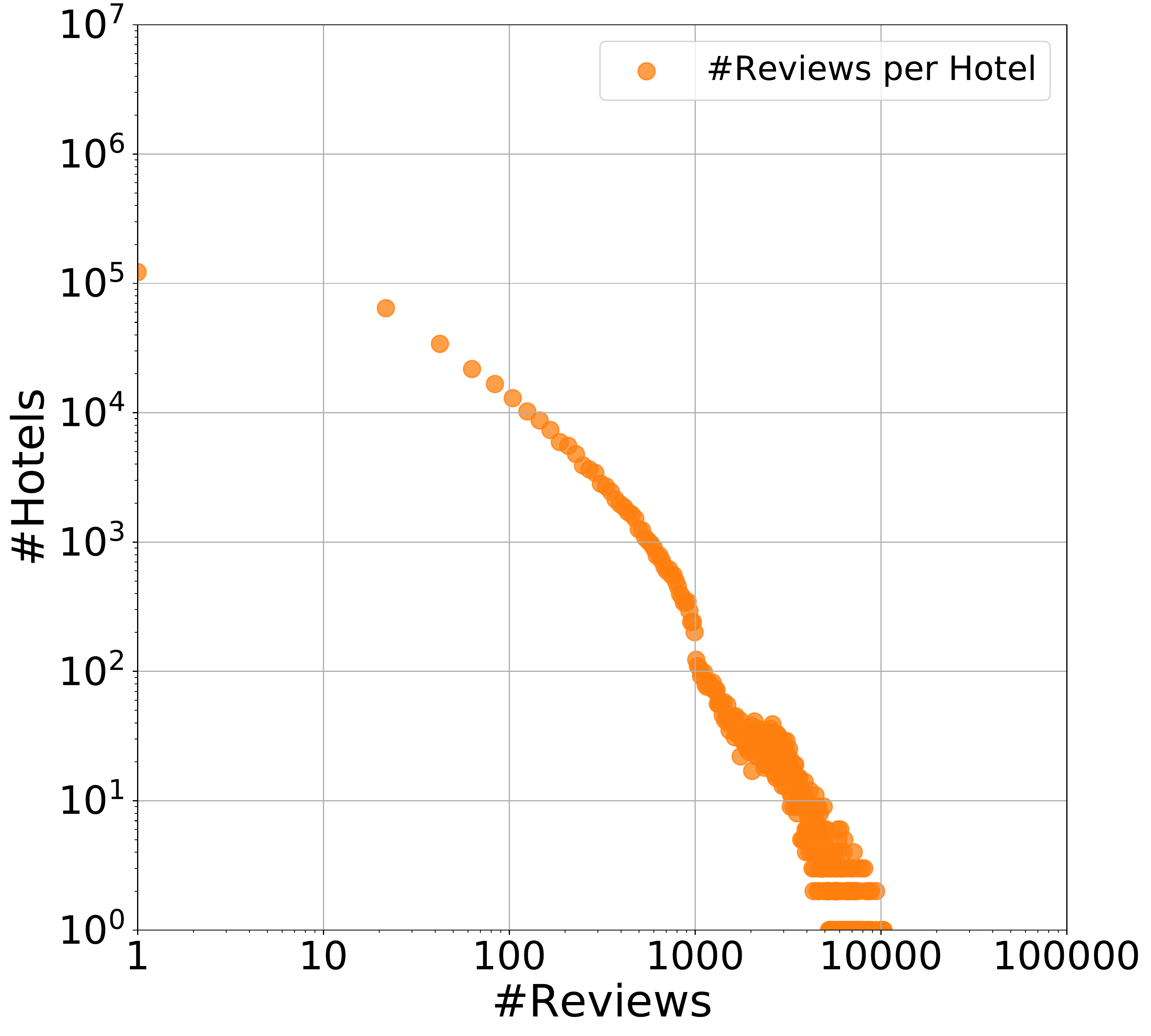}
  \caption{\label{fig_nb_reviews_item}Number of reviews per item. Mean: $137.69 \pm 342.94$.}
\end{subfigure}\hfil 

\medskip
\begin{subfigure}{0.45\textwidth}
  \includegraphics[width=\linewidth]{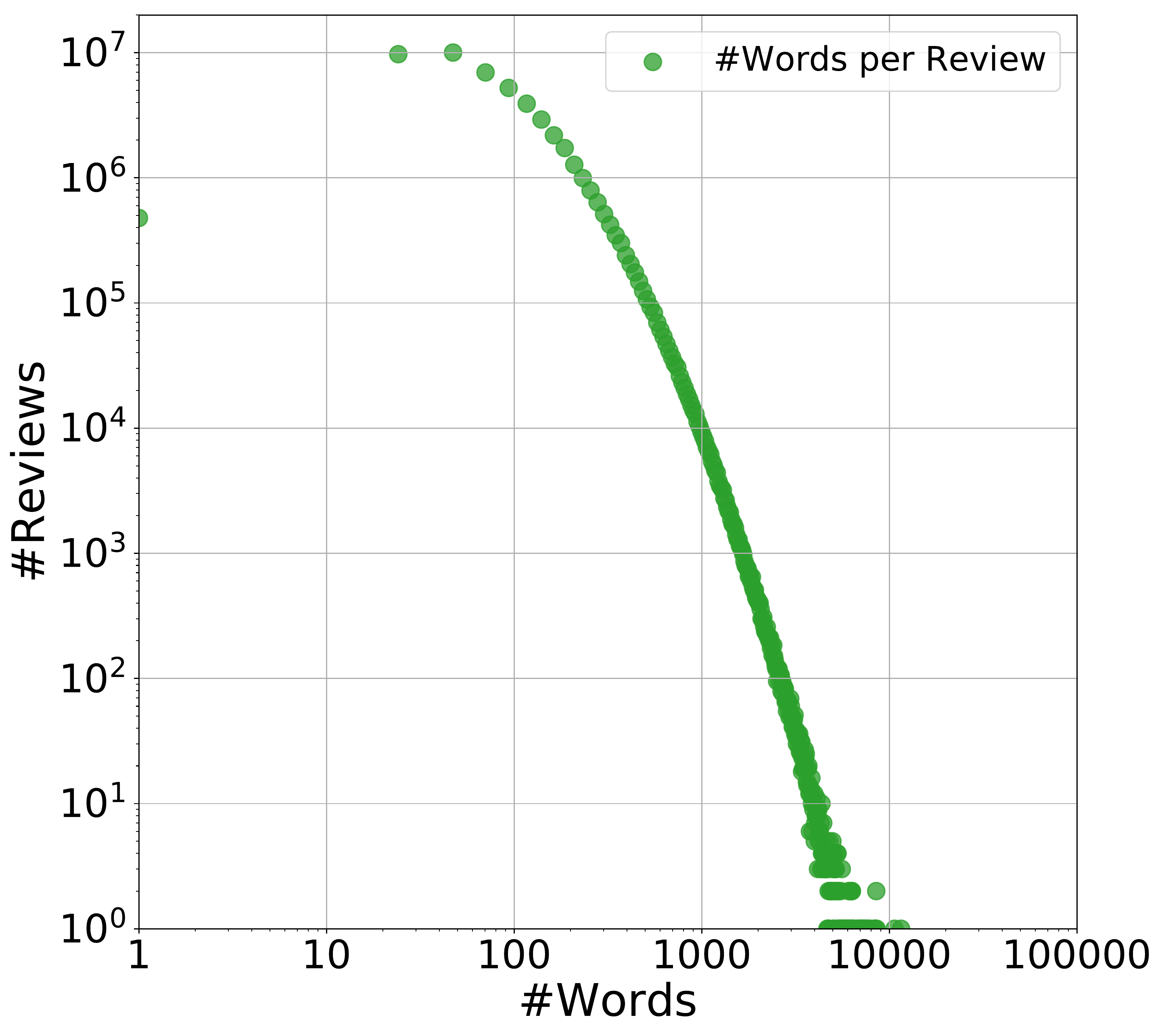}
  \caption{Number of words per review. Mean: $125.57 \pm 128.91$.}
  \label{fig_nb_words_review}
\end{subfigure}\hfil 
\begin{subfigure}{0.39\textwidth}
  \includegraphics[width=\linewidth]{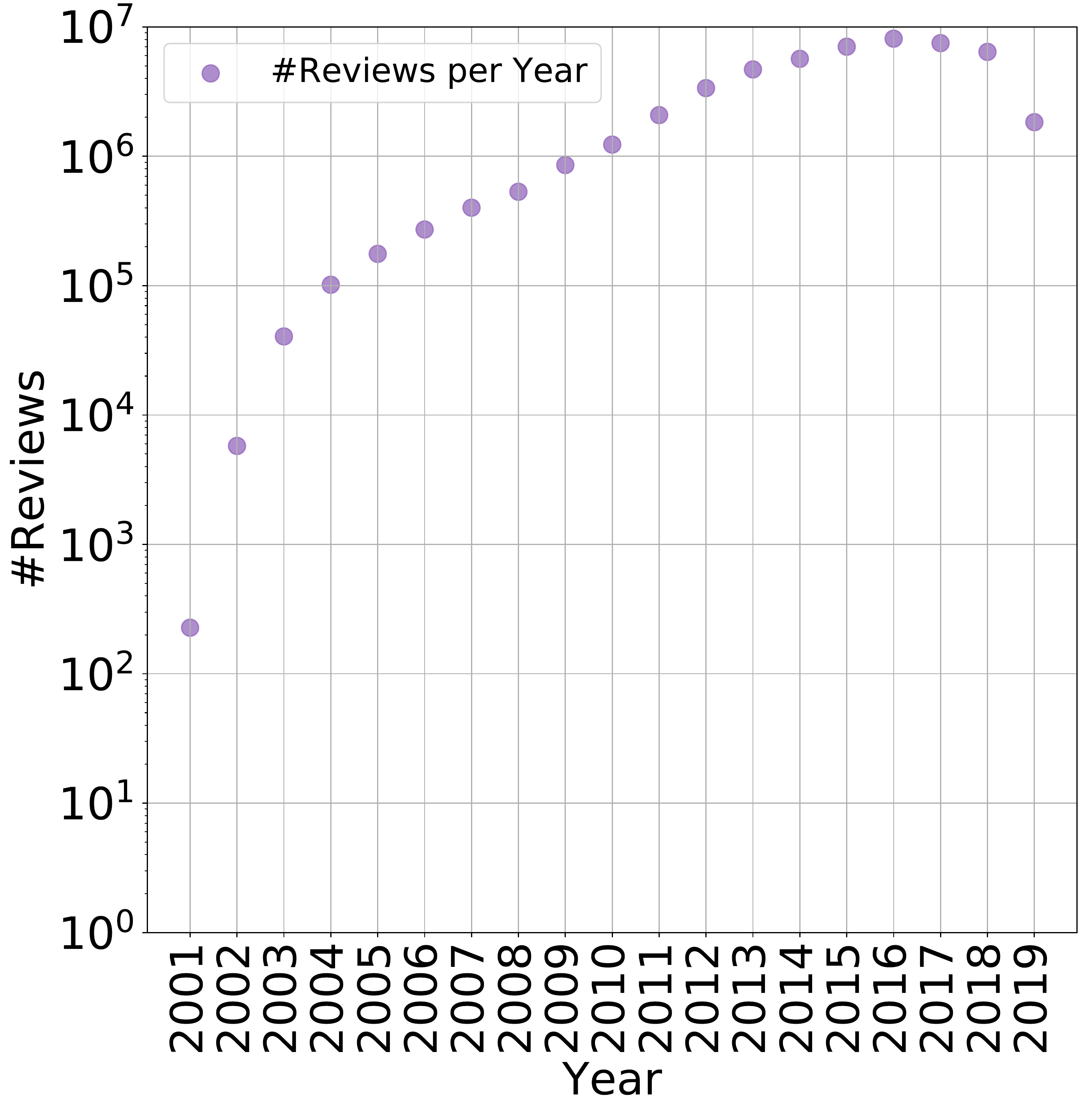}
  \caption{\label{fig_nb_reviews_year}Number of reviews per year.}
  \end{subfigure}\hfil 
    \caption{Histograms of multiple attributes of HotelRec, in logarithmic scales: number of reviews per user, item and year, and number of words per review.}
    \label{fig_histograms}
\end{figure*}

HotelRec includes $50\,264\,531$ hotel reviews from TripAdvisor in a period of nineteen years (from February $1$, $2001$ to May $14$, $2019$). The distribution of reviews over the years is available in Figure~\ref{fig_nb_reviews_year}. There is a significant activity increase of users from $2001$ to $2010$. After this period, the number of reviews per year grows slowly and oscillates between one to ten million. 

In total, there are $21\,891\,294$ users. The distribution of reviews per user is shown in Figure~\ref{fig_nb_reviews_user}. Similarly to other recommender datasets \cite{he2016ups,meyffret:hal-01010246}, the distribution resembles a Power-law distribution: many users write one or a few reviews. In HotelRec, $67.55\%$ users have written only one review, and $90.73\%$ with less than five reviews. Additionally, in the $5$-core subset, less than $15\%$ of $2\,012\,162$ users had a peer with whom they have co-rated three or more hotels. Finally, the average user has $2.24$ reviews, and the median is~$1.00$.

Relating to the items, there are $365\,056$ hotels, which is roughly $60$ times smaller than the number of users. This ratio is also consistent with other datasets \cite{mcauley2013hidden,McAuley2012}.

Figure~\ref{fig_nb_reviews_item} displays the distribution of reviews per hotel. The distribution also has a shape of a Power-law distribution, but its center is closer to $3\,000$ than the $100$ of the user distribution. However, in comparison, only $0.26\%$ hotels have less than five reviews and thus, the average reviews per hotel and the median are higher: $137.69$ and $41.00$.

Finally, we analyze the distribution of words per review, to understand how much people write about hotels. The distribution of words per review is shown in Figure~\ref{fig_nb_words_review}. The average review length is $125.57$ words, which is consistent with other studies \cite{mcauley2013hidden}.

\subsection{Overall and Sub-Ratings}
\label{sec_asp_corr}

When writing a review, the \textit{Overall} rating is mandatory: it represents the evaluation of the whole user experience towards a hotel. It is consequently available for all reviews in HotelRec. However, sub-ratings only assess one or more particular aspects (up to eight), such as \textit{Service}, \textit{Cleanliness}, or \textit{Location}. Additionally, they are optional: the user can choose how many and what aspects to evaluate. Among all the reviews, $35\,836\,414$ ($71.30\%$) have one or several sub-ratings, with a maximum of eight aspects. The distribution of the number of assessed fine-grained aspects is shown in Table~\ref{nb_subrating_descr}, where \textit{All} represents the coverage over the whole set of reviews, and \textit{With Sub-Ratings} over the set of reviews having sub-ratings (i.e., approximately $35$ million). Interestingly, most of the sub-ratings are evaluated in a group of three or six aspects. We hypothesize that this phenomenon came from a limitation of TripAdvisor on the user interface, where the set of aspects to evaluate was predefined.

\begin{table}[h]
  \centering
  \begin{tabular}{cc@{}c}
    & \multicolumn{2}{c}{\textbf{Coverage (\%)}}\\
    \textbf{\# Rated} & \multicolumn{1}{c}{\multirow{2}{*}{{\rotatebox[origin=c]{0}{\centering All}}}} & \multicolumn{1}{c}{\multirow{1}{*}{{\rotatebox[origin=c]{0}{\centering With Sub-}}}}\\
    \textbf{Aspects} &  & \multicolumn{1}{c}{\multirow{1}{*}{{\rotatebox[origin=c]{0}{\centering Ratings}}}} \\
        \hline
$1$ & \ $  0.404$ &  \ $ 0.566$ \\
$2$ & \ $  0.893$ &  \ $ 1.252$ \\
$3$ & $ 29.474$ &  $41.341$ \\
$4$ & \ $  2.220$ &  \ $ 3.113$ \\
$5$ & \ $  5.201$ &  \ $ 7.295$ \\
$6$ & $ 31.982$ &  $44.858$ \\
$7$ & \ $  1.120$ &  \ $ 1.572$ \\
$8$ & \ $  0.002$ &  \ $ 0.002$ \\
    \end{tabular}
  \caption{\label{nb_subrating_descr}Statistics of the number of rated fine-grained aspects in the HotelRec dataset. Coverage is the ratio of reviews having $i$~sub-ratings over: \textit{All} reviews, and \textit{only} reviews \textit{With Sub-Ratings} available.}
\end{table}

We analyze in Table~\ref{rating_subrating_descr} the distribution of the reviews with fine-grained and \textit{Overall} ratings. Unsurprisingly, the \textit{Overall} rating is always available as it is mandatory. In terms of aspects, there is a group of six that are majorly predominant (following the observation in Table~\ref{nb_subrating_descr}), and two that are rarely rated: \textit{Check-In} and \textit{Business Service}. Surprisingly, these two aspects are not sharing similar rating averages and percentiles than the others. We explain this difference due to the small number of reviews rating them (approximately $2\%$). Furthermore, most ratings across aspects are positive: the $25^{th}$ percentile is $4$, with an average of $4.23$ and a median of $5$.

\begin{table}[h]
  \centering
  \begin{tabular}{@{}l@{}rc@{\hspace*{1mm}}c@{\hspace*{1mm}}c@{\hspace*{1mm}}c@{\hspace*{1mm}}}
    \textbf{Aspect} & Coverage (\%) & Average & $25^{th}$ & $50^{th}$ & $75^{th}$ \\
        \hline
Overall & $100.00$& $4.15 \pm 1.12$ &  $4$ & $5$ & $5$\\
Service & $ 99.27$& $4.29 \pm 1.09$ &  $4$ & $5$ & $5$\\
Check-In & $  2.73$& $4.00 \pm 1.20$ &  $3$ & $4$ & $5$\\
Business Serv. & $  1.69$& $3.65 \pm 1.25$ &  $3$ & $4$ & $5$\\
Location & $ 71.22$& $4.40 \pm 0.88$ &  $4$ & $5$ & $5$\\
Value & $ 73.63$& $4.12 \pm 1.13$ &  $4$ & $5$ & $5$\\
Cleanliness & $ 73.69$& $4.33 \pm 1.03$ &  $4$ & $5$ & $5$\\
Rooms & $ 70.92$& $4.12 \pm 1.10$ &  $4$ & $4$ & $5$\\
Sleep Quality & $ 63.04$ & $4.21 \pm 1.08$ &  $4$ & $5$ & $5$\\
    \end{tabular}
  \caption{\label{rating_subrating_descr}Descriptive statistics of the ratings of the \textit{Overall} and fine-grained aspect ratings (e.g., \textit{Service}, \textit{Rooms}). Coverage describes the ratio of reviews having a particular fine-grained rating. The other columns represent the average, and the $25^{th}$, $50^{th}$ (median), $75^{th}$ percentiles of the individual ratings.}
\end{table}

\begin{figure}
    \centering
    \includegraphics[width=\linewidth]{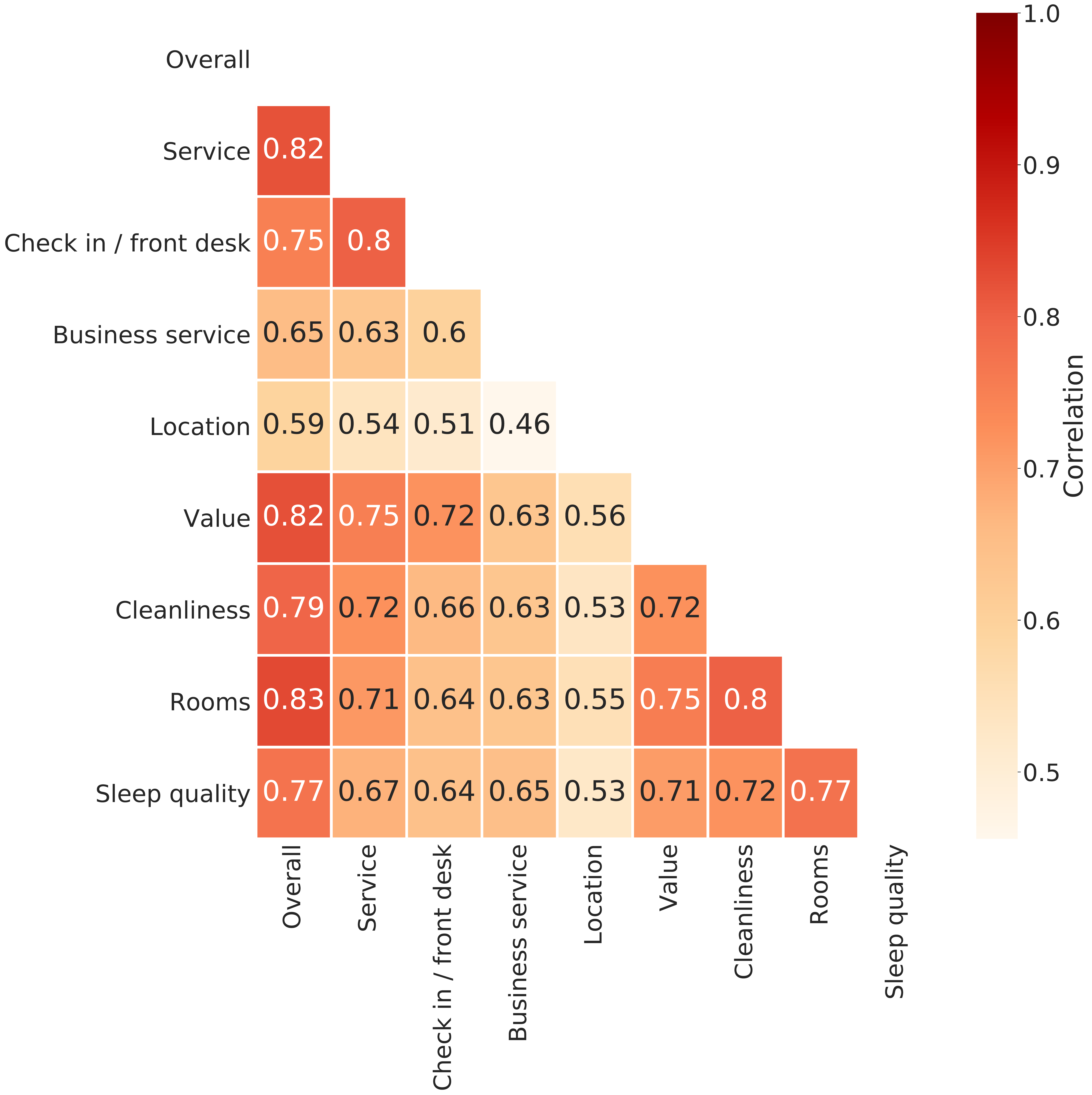}
    \caption{\label{fig_corr}Pearson correlation between all fine-grained and overall ratings. All aspect pairs are highly correlated. }
    
\end{figure}

Finally, in Figure~\ref{fig_corr}, we computed the Pearson correlation of ratings between all pairs of aspects, including fine-grained and \textit{Overall} ones. Interesting, all aspect-pairs have a correlation between $0.46$ and $0.83$. We observe that \textit{Service}, \textit{Value}, and \textit{Rooms} correlate the most with the \textit{Overall} ratings. Unsurprisingly, the aspect pair \textit{Service}-\textit{Check In} and \textit{Rooms}-\textit{Cleanliness} have a correlation of $0.80$, because people often evaluate them together in a similar fashion. Interestingly, \textit{Location} is the aspect that correlates the least with the others, followed by \textit{Business Service}, and \textit{Check-In}.

\section{Experiments and Results}

In this section, we first describe two different $k$-core subsets of the HotelRec dataset that we used to evaluate multiple baselines on two tasks: rating prediction and recommendation performance. We then detail the models we employed, and discuss their results.

\subsection{Datasets}

We used the aforementioned dataset HotelRec, containing approximately $50$ million hotel reviews. The characteristics of this dataset are described in Section~\ref{sec_descr} and Section~\ref{sec_asp_corr}
Following the literature \cite{he2017neural,cheng2018aspect}, we focused our evaluation on two $k$-core subsets of HotelRec, with at least $k$~reviews for each user or item. In this paper, we employed the most common values for $k$: $5$ and $20$. We randomly divided each of the datasets into $80/10/10$ for training, validation, and testing subsets.

From each review, we kept the corresponding "userID", "itemID", rating (from $1$ to $5$ stars), written text, and date. We preprocessed the text by lowering and tokenizing it. Statistics of both subsets are shown in Table~\ref{stat_core_dataset}.

\subsection{Evaluation Metrics and Baselines}

We evaluated different models on the HotelRec subsets, $5$-core and $20$-core, on two tasks: rating prediction and recommendation performance. We have separated the evaluation because most models are only tailored for one of the tasks but not both. Therefore, we applied different models for each task and evaluated them separately.

For the rating prediction task, following the literature, we reported the results in terms of Mean Square Error (MSE) and Root Mean Square Error (RMSE). We assessed the recommendation performance of a ranked list by Hit Ratio (HR) and Normalized Discounted Cumulative Gain (NDCG) \cite{he2015trirank}, as in \newcite{he2017neural}. We truncated the ranked list at $5$, $10$ and $20$. The HR measures whether a new item is on the top-$k$ list and NDCG measures the position of the hit by assigning higher scores to hits at top ranks. As in \newcite{he2017neural}, we computed both metrics for each test user and reported the average score. Regarding the models, we employed the following baselines:
\begin{itemize}
    \item \textbf{Mean}: A simple model that predicts a rating by the mean ratings of the desired item. It is a good baseline in recommendation~\cite{musat2015personalizing};
    \item \textbf{HFT}~\cite{mcauley2013hidden}: A latent-factor approach combined with a topic model that aims to find topics in the review text that correlate with latent factors of the users and the items;
    \item \textbf{TransNet(-Ext)}: The model is based on \newcite{zheng2017joint}, which learns a user and item profile based on former reviews using convolutional neural networks, and predicts the ratings using matrix factorization methods afterward. They added a regularizer network to improve performance. TransNet-Ext is an extension of TransNet by using a collaborative-filtering component in addition to user and item reviews history.
\end{itemize}

For the recommendation performance task, we used the following models :
\begin{itemize}
    \item \textbf{RAND}: A simple model recommending random items;
    \item \textbf{POP}~\cite{rendle2009bpr}: Another non-personalized recommender method, where items are recommended based on their popularity (i.e., the number of interactions with users). It is a common baseline to benchmark the recommendation performance;
    \item \textbf{ItemKNN/UserKNN}~\cite{Sarwar2001}: Two standard item-based (respectively user-based) collaborative filtering methods, using $k$ nearest neighbors;
    \item \textbf{PureSVD}~\cite{cremonesi2010performance}: A similarity based approach that constructs a similarity matrix through the SVD decomposition of the rating matrix; 
    \item \textbf{GMF}~\cite{he2017neural}: A generalization of the matrix factorization method that applies a linear kernel to model the latent feature interactions;
    \item \textbf{MLP}~\cite{he2017neural}: Similar than GMF, but it models the interaction of latent features with a neural network instead of a linear kernel;
    \item \textbf{NeuMF}~\cite{he2017neural}: A model combining GMF and MLP to better model the complex user-item interactions.
\end{itemize}

Due to the large size of the HotelRec dataset, especially in the $5$-core setting (around $20$ million reviews), running an extensive hyper-parameter tuning for each \textit{neural} model would require a high time and resource budget. Therefore, for the neural model, we used the default parameters from the original implementation and a random search of three trials. For all other models (i.e., HFT, ItemKNN, UserKNN, PureSVD), we ran a standard grid search over the parameter sets.

\subsection{Rating Prediction}

\begin{table}[h]
  \centering
  \begin{tabular}{l@{}lcc}
      & \textbf{Models} & MSE & RMSE \\
        \hline
      \multicolumn{1}{c}{\multirow{5}{*}{{\rotatebox[origin=c]{90}{\centering \textit{5-core}}}}}  & Mean & $0.7769$ & $0.8814$ \\
        & HFT - $K=5$ & $0.7667$ & $0.8756$\\
        & HFT - $K=10$ & $\mathbf{0.7520}$ & $\mathbf{0.8672}$\\
        & TransNet & $0.8003$ & $0.8946$\\
        & TransNet-Ext & $0.8319$ & $0.9121$ \\
        \hline
        \multicolumn{1}{c}{\multirow{5}{*}{{\rotatebox[origin=c]{90}{\centering \textit{20-core}}}}} 
        & Mean & $\mathbf{0.7546}$ & $\mathbf{0.8687}$\\
        & HFT - $K=5$ & $0.7720$ & $0.8786$ \\
        & HFT - $K=10$ & $0.7872$ & $0.8872$\\
        & TransNet & $0.8394$ & $0.9162$ \\
        & TransNet-Ext & $1.0754$ & $1.0370$ \\
    \end{tabular}
  \caption{\label{perf_rating}Evaluation of rating prediction in terms of Mean Square Error (MSE) and Root Mean Square Error (RMSE).}
\end{table}

\begin{table*}[h]
  \centering
  \begin{tabular}{llcccccc}
    & \textbf{Models} & HR@5 & NDCG@5 & HR@10 & NDCG@10 & HR@20 & NDCG@20\\
        \hline
    \multicolumn{1}{c}{\multirow{8}{*}{{\rotatebox[origin=c]{90}{\centering \textit{5-core}}}}} & RAND & $0.0000$ & $0.0000$ & $0.0000$ & $0.0000$ & $0.0001$ & $0.0000$\\
    & POP & $0.0018$ & $0.0007$ & $0.0034$ & $0.0010$ & $0.0060$ & $0.0014$ \\
    & ItemKNN & $0.0162$ & $0.0072$ & $0.0238$ & $0.0088$ & $0.0340$ & $0.0103$ \\
    & UserKNN & $0.0118$ & $0.0053$ & $0.0176$ & $0.0065$ & $0.0248$ & $0.0077$ \\
    & PureSVD & $0.0089$ & $0.0039$ & $0.0141$ & $0.0050$ & $0.0221$ & $0.0064$\\
    & GMF & $0.3899$ & $0.2761$ & $0.5340$ & $0.3237$ & $0.7055$ & $0.3666$ \\
    & MLP & $0.4320$ & $0.3070$ & $0.5734$ & $0.3533$ & $0.7251$ & $0.3915$\\
    & NeuMF & $\mathbf{0.4981}$ & $\mathbf{0.3589}$ & $\mathbf{0.6481}$ & $\mathbf{0.4066}$ & $\mathbf{0.7836}$ & $\mathbf{0.4401}$\\
    \hline
    \multicolumn{1}{c}{\multirow{8}{*}{{\rotatebox[origin=c]{90}{\centering \textit{20-core}}}}} & RAND & $0.0004$ & $0.0000$ & $0.0010$ & $0.0002$ & $0.0019$ & $0.0002$ \\
    & POP & $0.0064$ & $0.0016$ & $0.0109$ & $0.0021$ & $0.0210$ & $0.0030$  \\
    & ItemKNN & $0.0236$ & $0.0061$ & $0.0411$ & $0.0084$ & $0.0682$ & $0.0110$ \\
    & UserKNN & $0.0208$ & $0.0054$ & $0.0360$ & $0.0073$ & $0.0587$ & $0.0095$\\
    & PureSVD & $0.0216$ & $0.0055$ & $0.0375$ & $0.0075$ & $0.0616$ & $0.0099$\\
    & GMF & $0.3705$ & $0.2565$ & $0.5219$ & $0.3047$ & $0.6913$ & $0.3477$\\
    & MLP & $0.3731$ & $0.2564$ & $0.5251$ & $0.3050$ & $0.6962$ & $0.3496$ \\
    & NeuMF & $\mathbf{0.4274}$ & $\mathbf{0.3000}$ & $\mathbf{0.5776}$ & $\mathbf{0.3483}$ & $\mathbf{0.7354}$ & $\mathbf{0.3884}$\\
    \end{tabular}
  \caption{\label{perf_rec}Evaluation of Top-$K$ recommendation performance in terms of Hit Ratio (HR) and Normalized Discounted Cumulative Gain (NDCG).}
\end{table*}

We show in Table~\ref{perf_rating} the performance in terms of the mean square error (MSE) and the root mean square error (RMSE). Surprisingly, we observe that the neural network TransNet and its extension perform poorly in comparison to the matrix factorization model HFT and the simple Mean baselines. Although TransNet learns a user and item profile based on the most recent reviews, it cannot capture efficiently the interaction from these profiles. Moreover, the additional collaborative-filtering component in TransNet-Ext seems to worsen the performance, which is consistent with the results of \newcite{musat2013recommendation}; in the hotel domain, the set users who have rated the same hotels is sparser than usual recommendation datasets.

Interestingly, the Mean model obtains the best performance on the $20$-core subset, while HFT achieves the best performance on the $5$-core subset. We hypothesize that HFT and TransNet(-Ext) models perform better on the $5$-core than $20$-core subset, because of the number of data. More specifically, HFT employs Latent Dirichlet Allocation~\cite{Blei2003} to approximate topic and word distributions. Thus, the probabilities are more accurate with a text corpus approximately ten times larger.

\subsection{Recommendation Performance}

The results of the baselines are available in Table~\ref{perf_rec}. HR@$k$ and NDCG@$k$ correspond to the Hit Ratio (HR) and Normalized Discounted Cumulative Gain (NDCG), evaluated on the top-$k$ computed ranked items for a particular test user, and then averaged over all test users.

First, we can see that NeuMF significantly outperforms all other baselines on both $k$-core subsets. The other methods GMF and MLP - both used within NeuMF - also show quite strong performance and comparable performance. However, NeuFM achieves higher results by fusing GMF and MNLP within the same model. Second, if we compare ItemKNN and UserKNN, we observe that on both subsets, the user collaborative filtering approach underperform compared to its item-based variant, that matches the founding in the rating prediction task of the previous section, and the work of \newcite{musat2013recommendation,musat2015personalizing}. Additionally, PureSVD achieves comparable results with UserKNN.

Finally, the two non-personalized baselines RAND and POP obtain unsurprisingly low results, indicating the necessity of modeling user's preferences to a personalized recommendation.

\section{Conclusion}

In this work, we introduce HotelRec, a novel large-scale dataset of hotel reviews based on TripAdvisor, and containing approximately $50$ million reviews. Each review includes the user profile, the hotel URL, the overall rating, the summary, the user-written text, the date, and multiple sub-ratings of aspects when provided. To the best of our knowledge, HotelRec is the largest publicly available dataset in the hotel domain ($50M$ versus $0.9M$) and additionally, the largest recommendation dataset in \textit{a single domain} and with \textit{textual reviews} ($50M$ versus $22M$).

We further analyze the HotelRec dataset and provide benchmark results for two tasks: rating prediction and recommendation performance. We apply multiple common baselines, from non-personalized methods to competitive models, and show that reasonable performance could~be obtained, but still far from results achieved in other domains in the literature. 

In future work, we could easily increase the dataset with other languages and use it for multilingual recommendation. We release HotelRec for further research: \url{https://github.com/Diego999/HotelRec}.

\section{Bibliographical References}
\label{main:ref}

\bibliographystyle{lrec}

\end{document}